\documentclass{article}
\usepackage{graphicx}
\usepackage{amsmath}

\begin{document}

\textbf{Lorentz and ``apparent'' transformations of the electric and }

\textbf{magnetic fields}\bigskip \medskip

\qquad Tomislav Ivezi\'{c}

\qquad\textit{Ru%
\mbox
{\it{d}\hspace{-.15em}\rule[1.25ex]{.2em}{.04ex}\hspace{-.05em}}er Bo\v
{s}kovi\'{c} Institute, P.O.B. 180, 10002 Zagreb, Croatia}

\textit{\qquad ivezic@irb.hr\bigskip \bigskip }

\noindent It is recently discovered that the usual transformations of the
three-dimensional (3D) vectors of the electric and magnetic fields differ
from the Lorentz transformations (LT) (boosts) of the corresponding 4D
quantities that represent the electric and magnetic fields. In this paper,
using geometric algebra formalism, this fundamental difference is examined
representing the electric and magnetic fields by bivectors. \medskip \bigskip

\noindent \textbf{I. INTRODUCTION\bigskip }

Recently$^{1,2,3}$ it is proved that, contrary to the general belief, the
usual transformations of the three-dimensional (3D) vectors of the electric
and magnetic fields, see, e.g., Ref. 4, Eqs. (11.148) and (11.149), differ
from the Lorentz transformations (LT) (boosts) of the corresponding 4D
quantities that represent the electric and magnetic fields. (The usual
transformations will be called the ``apparent'' transformations (AT) and the
name will be explained in Sec. IV.) Comparisons with experiments, the
motional emf,$^{2}$ the Faraday disk$^{3}$ and the Trouton-Noble experiment,$%
^{5,6}$ show that this new approach$^{1-3,5.6}$ with 4D geometric quantities
always agrees with the principle of relativity and with experiments, while
it it is not the case with the usual approach in which the electric and
magnetic fields are represented by the 3D vectors $\mathbf{E}$ and $\mathbf{B%
}$ that transform according to the AT. The mentioned agreement with
experiments is independent of the chosen reference frame and of the chosen
system of coordinates in it. The main point in the geometric approach$%
^{1-3,5.6}$ is that the physical meaning, both theoretically and \emph{%
experimentally}, is attributed to 4D geometric quantities, and not, as
usual, to 3D quantities.

In this paper I shall present a simplified version of the proof of the
difference between the LT and the AT that is already given in Secs. 3.3 and
4 in Ref. 2. For all mathematical details for the used geometric algebra
formalism readers can consult Refs. 7, 8.

As shown in Refs. 2, 3 the electric and magnetic fields can be represented
by different algebraic objects; 1-vectors, bivectors or their combination.
The representation with 1-vectors $E$ and $B$ is simpler than others and
also closer to the usual expressions with the 3D vectors $\mathbf{E}$ and $%
\mathbf{B}$, but here we shall only deal with bivectors. The reason is that
the representation with bivectors, as in our Eq. (\ref{fh}), is always
employed in Refs. 7, 8 and we want to make comparison with their results. In
Sec. II. A a new Lorentz invariant representation, $E_{v}$ and $B_{v}$, is
presented that is introduced in Refs. 2 and 3. In Sec. II. B we simply
derive from $E_{v}$ and $B_{v}$ the observer dependent expressions for the
electric $\mathbf{E}_{H}$ and magnetic $\mathbf{B}_{H}$ fields, which are
always exploited in Refs. 7, 8. $\mathbf{E}_{H}^{\prime }$ (and $\mathbf{B}%
_{H}^{\prime }$), which are the LT (the active ones) of $\mathbf{E}_{H}$
(and $\mathbf{B}_{H}$)$,$ Eqs. (\ref{ch}) and (\ref{eh}), are derived in
Sec. III using the fact that every multivector must transform under the
active LT in the same way, i.e., according to Eq. (\ref{bo}). Furthermore,
it is known that any multivector, when written in terms of components and a
basis, must remain unchanged under the passive LT, like some general
bivector in Eq. (\ref{enc}). Hence observers in relatively moving inertial
frames $S$ and $S^{\prime }$ ``see'' the same $\mathbf{E}_{H}$, i.e., Eq. (%
\ref{n1}) holds for $\mathbf{E}_{H}$. These fundamental achievements for the
LT of bivectors $\mathbf{E}_{H}$ (and $\mathbf{B}_{H}$) are first obtained
in Ref. 2. Hestenes$^{7}$ and the Cambridge group$^{8}$ derived the
transformations for $\mathbf{E}_{H}$ and $\mathbf{B}_{H}$ (Ref. 7, \textit{%
Space-Time Algebra,} Eq. (18.22), \textit{New Foundations for Classical
Mechanics,} Ch. 9, Eqs. (3.51a,b) and Ref. 8, Sec. 7.1.2, Eq. (7.33)) in the
way that is presented in Sec. IV, Eqs. (\ref{boc}) and (\ref{es}) for $%
\mathbf{E}_{H,at}^{\prime }$, and Eqs. (\ref{et}), (\ref{is}) and (\ref{ib})
for components. The transformations for components, Eqs. (\ref{is}) and (\ref
{ib}), are identical to the usual transformations for components of the 3D $%
\mathbf{E}$ and $\mathbf{B}$, Ref. 4, Eq. (11.148). Such usual
transformations are quoted in every textbook and paper on relativistic
electrodynamics already from the time of Einstein's fundamental paper.$^{9}$
They are always considered (including Refs. 7, 8) to be the LT of the
electric and magnetic fields. However it is obvious from (\ref{boc}) and (%
\ref{es}) that $\mathbf{E}_{H,at}^{\prime }$ is not obtained by the active
LT from $\mathbf{E}_{H}$, since (\ref{boc}) is drastically different than
the correct LT (\ref{bo}) and (\ref{ch}). Furthermore, as seen from Eq. (\ref
{n2}), Eq. (\ref{enc}) is not fulfilled, which means that $\mathbf{E}_{H}$
and $\mathbf{E}_{H,at}^{\prime }$ are not the same physical quantity for
relatively moving observers in $S$ and $S^{\prime }$. Again, completely
different result than that one obtained by the correct passive LT, Eq. (\ref
{n1}). This shows that neither the usual transformations of the electric and
magnetic fields from Refs. 7, 8 nor the usual transformations for
components, Ref. 4, Eqs. (11.148), are the LT. The conclusions are given in
Sec. V together with the short presentation of the fundamental difference
between the LT and the AT when dealing with 1-vectors $E$ and $B$. \bigskip
\medskip

\noindent \textbf{II. ELECTRIC\ AND\ MAGNETIC\ FIELDS AS\ BIVECTORS \medskip
}

In this geometric approach physical quantities will be represented by 4D
geometric quantities, multivectors, that are defined without reference
frames, or, when some basis has been introduced, these quantities are
represented as 4D geometric quantities comprising both components and a
basis. For simplicity and for easier understanding, only the standard basis
\{$\gamma _{\mu };\ 0,1,2,3$\} of orthonormal 1-vectors, with timelike
vector $\gamma _{0}$ in the forward light cone, will be used in the
Minkowski spacetime $M^{4}$, but remembering that the approach with 4D
geometric quantities holds for any choice of basis in $M^{4}$. The basis
vectors $\gamma _{\mu }$ generate by multiplication a complete basis for the
spacetime algebra: $1,$ $\gamma _{\mu },$ $\gamma _{\mu }\wedge \gamma _{\nu
},$ $\gamma _{\mu }\gamma _{5,}$ $\gamma _{5}$ (16 independent elements). $%
\gamma _{5}$ is the right-handed unit pseudoscalar, $\gamma _{5}=\gamma
_{0}\wedge \gamma _{1}\wedge \gamma _{2}\wedge \gamma _{3}$. Any multivector
can be expressed as a linear combination of these 16 basis elements of the
spacetime algebra. It is worth noting that the standard basis \{$\gamma
_{\mu }$\} corresponds, in fact, to the specific system of coordinates,
i.e., to Einstein's system of coordinates. In the Einstein system of
coordinates the Einstein synchronization$^{9}$ of distant clocks and
Cartesian space coordinates $x^{i}$ are used in the chosen inertial frame of
reference. However different systems of coordinates of an inertial frame of
reference are allowed and they are all equivalent in the description of
physical phenomena.\bigskip \medskip

\noindent \textbf{A. Lorentz invariant electric and magnetic fields}\bigskip

The electromagnetic field is represented by a bivector-valued function $%
F=F(x)$ on the spacetime. As shown in Refs. 2, 3 the observer independent $F$
can be decomposed into two bivectors $E_{v}$ and $B_{v}$ representing the
electric and magnetic fields and the unit time-like 1-vector $v/c$ as
\begin{align}
F& =E_{v}+cIB_{v}\mathbf{,\quad }E_{v}=(1/c^{2})(F\cdot v)\wedge
v=(1/2c^{2})(F-vFv),  \notag \\
IB_{v}& =(1/c^{3})(F\wedge v)\cdot v=(1/2c^{3})(F+vFv),  \label{he}
\end{align}
where $I$ is the unit pseudoscalar and $v$ is the velocity (1-vector) of a
family of observers who measures $E_{v}$ and $B_{v}$ fields. Observe that $%
E_{v}$ and $B_{v}$ depend not only on $F$ but on $v$ as well. All quantities
$F$, $E_{v}$, $B_{v}$, $I$ and $v$ are defined without reference frames. ($I$
is defined algebraically without introducing any reference frame, as in Ref.
10 Sec. 1.2.) Such 4D geometric quantities will be called the absolute
quantities (AQs), while their representations in some basis will be called
coordinate-based geometric quantities (CBGQs). For example, in the $\left\{
\gamma _{\mu }\right\} $ basis the AQ $E_{v}$ from (\ref{he}) is represented
by the following CBGQ $E_{v}=(1/c^{2})F^{\mu \nu }v_{\nu }v^{\beta }\gamma
_{\mu }\wedge \gamma _{\beta }$\bigskip \medskip

\noindent \textbf{B. Electric and magnetic fields in the}\textit{\ }$\gamma
_{0}$\textit{\ }- \textbf{frame\bigskip }

For comparison with the usual treatments$^{7,8}$ let us choose the frame in
which the observers who measure $E_{v}$ and $B_{v}$ are at rest. For them $%
v=c\gamma _{0}$. This frame will be called the frame of ``fiducial''
observers or the $\gamma _{0}$ - frame. In that frame $E_{v}$ and $B_{v}$
from (\ref{he}) become the observer dependent ($\gamma _{0}$ - dependent) $%
\mathbf{E}_{H}$ and $\mathbf{B}_{H}$ and instead of Eq. (\ref{he}) we have
\begin{eqnarray}
F &=&\mathbf{E}_{H}+c\gamma _{5}\mathbf{B}_{H}\mathbf{,\quad E}_{H}=(F\cdot
\gamma _{0})\gamma _{0}=(1/2)(F-\gamma _{0}F\gamma _{0}),  \notag \\
\gamma _{5}\mathbf{B}_{H} &=&(1/c)(F\wedge \gamma _{0})\gamma
_{0}=(1/2c)(F+\gamma _{0}F\gamma _{0}).  \label{fh}
\end{eqnarray}
(The subscript $H$ is for ``Hestenes.'') $E_{v}$ and $B_{v}$ in the $\gamma
_{0}$ - frame are denoted as $\mathbf{E}_{H}$ and $\mathbf{B}_{H}$ since
they are identical to 4D quantities used by Hestenes$^{7}$ and the Cambridge
group$^{8}$ for the representation of the electric and magnetic fields. We
note that such procedure is never used by Hestenes$^{7}$ and the Cambridge
group$^{8}$ since they deal from the outset only with $\gamma _{0}$ and thus
with a space-time split in the $\gamma _{0}$ - frame, i.e., with the
relations (\ref{fh}). This shows that the space-time split and the
corresponding observer dependent form for the electric and magnetic fields, (%
\ref{fh}), which is always used in Refs. 7, 8, is simply obtained in our
approach going to the frame of the ''fiducial'' observers, i.e., replacing
some general velocity $v$ in (\ref{he}) by $c\gamma _{0}$.

$\mathbf{E}_{H}$ and $\mathbf{B}_{H}$ from (\ref{fh}) can be written as
CBGQs in the standard basis $\left\{ \gamma _{\mu }\right\} $. They are
\begin{equation}
\mathbf{E}_{H}=F^{i0}\gamma _{i}\wedge \gamma _{0},\quad \mathbf{B}%
_{H}=(1/2c)\varepsilon ^{kli0}F_{kl}\gamma _{i}\wedge \gamma _{0}.
\label{aj}
\end{equation}
It follows from (\ref{aj}) that the components of $\mathbf{E}_{H},$ $\mathbf{%
B}_{H}$ in the $\left\{ \gamma _{\mu }\right\} $ basis (i.e., in the
Einstein system of coordinates) give rise to the tensor (components)$\;(%
\mathbf{E}_{H})^{\mu \nu }=\gamma ^{\nu }\cdot (\gamma ^{\mu }\cdot \mathbf{E%
}_{H})=(\gamma ^{\nu }\wedge \gamma ^{\mu })\cdot \mathbf{E}_{H},$ (and the
same for $(\mathbf{B}_{H})^{\mu \nu }$) which, written out as a matrix, have
entries
\begin{align}
(\mathbf{E}_{H})^{i0}& =F^{i0}=E^{i},\quad (\mathbf{E}_{H})^{ij}=0,  \notag
\\
(\mathbf{B}_{H})^{i0}& =(1/2c)\varepsilon ^{kli0}F_{kl}=B^{i},\quad (\mathbf{%
B}_{H})^{ij}=0.  \label{ad}
\end{align}
$(\mathbf{E}_{H})^{\mu \nu }$ is antisymmetric, i.e., $(\mathbf{E}_{H})^{\nu
\mu }=-(\mathbf{E}_{H})^{\mu \nu }$, and the same holds for $(\mathbf{B}%
_{H})^{\mu \nu }$. $(\mathbf{E}_{H})^{\mu \nu }$ from Eq. (\ref{ad}) can be
written in a matrix form as
\begin{equation}
(\mathbf{E}_{H})^{\mu \nu }=\left[
\begin{array}{cccc}
0 & -E^{1} & -E^{2} & -E^{3} \\
E^{1}=F^{10} & 0 & 0 & 0 \\
E^{2}=F^{20} & 0 & 0 & 0 \\
E^{3}=F^{30} & 0 & 0 & 0
\end{array}
\right] ,  \label{em}
\end{equation}
and readers can check that the same matrix form is obtained for $(\mathbf{B}%
_{H})^{\mu \nu }$. ($(\mathbf{B}_{H})^{10}=(1/c)F^{32}=B^{1}$.)

Thus we see from (\ref{aj}), and (\ref{ad}) or (\ref{em}), that

(i) both bivectors $\mathbf{E}_{H}$ and $\mathbf{B}_{H}$ are parallel to $%
\gamma _{0}$, $\mathbf{E}_{H}\wedge \gamma _{0}=\mathbf{B}_{H}\wedge \gamma
_{0}=0$, and consequently all space-space components of $(\mathbf{E}%
_{H})^{\mu \nu }$ and $(\mathbf{B}_{H})^{\mu \nu }$ are zero, $(\mathbf{E}%
_{H})^{ij}=(\mathbf{B}_{H})^{ij}=0$.

In the usual covariant approaches$^{4}$ the components of the 3D $\mathbf{E}$
and $\mathbf{B}$ are identified with six independent components of $F^{\mu
\nu }$ according to the relations

\begin{equation}
E_{i}=F^{i0},\quad B_{i}=(-1/2c)\varepsilon _{ikl}F_{kl}.  \label{sko1}
\end{equation}
In (\ref{sko1}) and hereafter the components of the 3D fields $\mathbf{E}$
and $\mathbf{B}$ are written with lowered (generic) subscripts, since they
are not the spatial components of the 4D quantities. This refers to the
third-rank antisymmetric $\varepsilon $ tensor too. The super- and
subscripts are used only on the components of the 4D quantities.

Comparing (\ref{ad}) and (\ref{sko1}) we see that they similarly identify
the components of the electric and magnetic fields with six independent
components of $F^{\mu \nu }$. However there are important differences
between the relations (\ref{aj}), (\ref{ad}) or (\ref{em}), and (\ref{sko1}%
). In the usual covariant approaches, e.g., Ref. 4, the 3D $\mathbf{E}$ and $%
\mathbf{B}$, as \emph{geometric quantities in the 3D space}, are constructed
from these six independent components of $F^{\mu \nu }$ and \emph{the unit
3D vectors }$\mathbf{i},$ $\mathbf{j},$ $\mathbf{k,}$ e.g., $\mathbf{E=}%
F^{10}\mathbf{i}+F^{20}\mathbf{j}+F^{30}\mathbf{k}$. Observe that the
mapping, i.e., the simple identification, Eq. (\ref{sko1}), of the
components $E_{i}$ and $B_{i}$ with some components of $F^{\mu \nu }$
(defined on the 4D spacetime) is not a permissible tensor operation, i.e.,
it is not a mathematically correct procedure. The same holds for the
construction of the \emph{3D vectors} $\mathbf{E}$ and $\mathbf{B}$ in which
the components of the \emph{4D quantity} $F^{\mu \nu }$ are multiplied with
\emph{the unit 3D vectors, }see Ref. 3 for the more detailed discussion. On
the other hand, as seen from Eqs. (\ref{aj}), (\ref{ad}) or (\ref{em}), $%
\mathbf{E}_{H}$ and $\mathbf{B}_{H}$ and their components $(\mathbf{E}%
_{H})^{\mu \nu }$ and $(\mathbf{B}_{H})^{\mu \nu }$ are obtained by a
correct mathematical procedure from the geometric 4D quantities $F$ and $%
\gamma ^{\mu }$. The components $(\mathbf{E}_{H})^{\mu \nu }$ and $(\mathbf{B%
}_{H})^{\mu \nu }$ are multiplied by the unit bivectors $\gamma _{i}\wedge
\gamma _{0}$ (4D quantities) to form the geometric 4D quantities $\mathbf{E}%
_{H}$ and $\mathbf{B}_{H}$. In such a treatment the unit 3D vectors\emph{\ }$%
\mathbf{i},$ $\mathbf{j},$ $\mathbf{k,}$ (geometric quantities in the \emph{%
3D space}) do not appear at any point.

Furthermore it is worth noting that $F^{\mu \nu }$ are only components
(numbers) that are (implicitly) determined in Einstein's system of
coordinates. Components are frame-dependent numbers (frame-dependent because
the basis refers to a specific frame). Components tell only part of the
story, while the basis contains the rest of the information about the
considered physical quantity. These facts are completely overlooked in all
usual covariant approaches and in the above identifications (\ref{sko1}) of $%
E_{i}$ and $B_{i}$ with some components of $F^{\mu \nu }$.\bigskip \medskip

\noindent \textbf{III. LT OF\ ELECTRIC\ AND\ MAGNETIC\ FIELDS AS\ BIVECTORS
\bigskip }

Let us now apply the active LT (only boosts are considered) to $\mathbf{E}%
_{H}$ and $\mathbf{B}_{H}$ from Eq. (\ref{aj}). In the usual geometric
algebra formalism$^{7,8}$ the LT are considered as active transformations;
the components of, e.g., some 1-vector relative to a given inertial frame of
reference (with the standard basis $\left\{ \gamma _{\mu }\right\} $) are
transformed into the components of a new 1-vector relative to the same frame
(the basis $\left\{ \gamma _{\mu }\right\} $ is not changed). Furthermore
the LT are described with rotors $R,$ $R\widetilde{R}=1,$ in the usual way
as $p\rightarrow p^{\prime }=Rp\widetilde{R}=p_{\mu }^{\prime }\gamma ^{\mu
}.$ Remember that the reverse $\widetilde{R}$ is defined by the operation of
reversion according to which $\widetilde{AB}=\widetilde{B}\widetilde{A},$ $%
\widetilde{a}=a,$ for any vector $a$, and it reverses the order of vectors
in any given expression. Every rotor in spacetime can be written in terms of
a bivector as $R=e^{\theta /2}.$ For boosts in arbitrary direction the rotor
$R$ is
\begin{equation}
R=e^{\theta /2}=(1+\gamma +\gamma \beta \gamma _{0}n)/(2(1+\gamma ))^{1/2},
\label{br}
\end{equation}
$\theta =\alpha \gamma _{0}n,$ $\beta $ is the scalar velocity in units of $%
c $, $\gamma =(1-\beta ^{2})^{-1/2}$, or in terms of an `angle' $\alpha $ we
have $\tanh \alpha =\beta ,$ $\cosh \alpha =\gamma ,$ $\sinh \alpha =\beta
\gamma ,$ and $n$ is not the basis vector but any unit space-like vector
orthogonal to $\gamma _{0};$ $e^{\theta }=\cosh \alpha +\gamma _{0}n\sinh
\alpha .$ One can also express the relationship between two relatively
moving frames $S$ and $S^{\prime }$ in terms of rotor as $\gamma _{\mu
}^{\prime }=R\gamma _{\mu }\widetilde{R}.$ For boosts in the direction $%
\gamma _{1}$ the rotor $R$ is given by the relation (\ref{br}) with $\gamma
_{1}$ replacing $n$ (all in the standard basis $\left\{ \gamma _{\mu
}\right\} $). For simplicity we shall only consider boosts in the direction $%
\gamma _{1}$.

As said in Sec. IV in Hestenes' paper$^{7}$ in AJP Lorentz rotations
preserve the geometric product. This implies that any multivector $M$
transforms by the active LT in the same way as mentioned above for the
1-vector $p$, i.e.,
\begin{equation}
M\rightarrow M^{\prime }=RM\widetilde{R},  \label{bo}
\end{equation}
see, e.g., Eq. (69) in Hestenes' paper$^{7}$ in AJP. It is not important is $%
M$ a simple blade or a Clifford aggregate, is it a function of some other
multivectors or it is not.

Hence, according to (\ref{bo}), under the active LT $\mathbf{E}_{H}$ from (%
\ref{fh}) must transform in the following way
\begin{equation}
\mathbf{E}_{H}^{\prime }=R[(1/2)(F-\gamma _{0}F\gamma _{0})]\widetilde{R}%
=(1/2)[F^{\prime }-\gamma _{0}^{\prime }F^{\prime }\gamma _{0}^{\prime
}]=(F^{\prime }\cdot \gamma _{0}^{\prime })\gamma _{0}^{\prime },  \label{ch}
\end{equation}
where $F^{\prime }=RF\widetilde{R}$ and $\gamma _{0}^{\prime }=R\gamma _{0}%
\widetilde{R}$. However, as will be shown in Sec. IV, it is surprising that
neither Hestenes$^{7}$ nor the Cambridge group$^{8}$ transform $\mathbf{E}%
_{H}$ in the way in which all other multivectors are transformed, i.e.,
according to (\ref{bo}) and (\ref{ch}).

When the active LT are applied to $\mathbf{E}_{H}$ from (\ref{aj}), thus
when $\mathbf{E}_{H}$ is written as a CBGQ, then $\mathbf{E}_{H}^{\prime }$
becomes
\begin{align}
\mathbf{E}_{H}^{\prime }& =R[E^{i}\gamma _{i}\wedge \gamma _{0}]\widetilde{R}%
=E^{1}\gamma _{1}\wedge \gamma _{0}+\gamma (E^{2}\gamma _{2}\wedge \gamma
_{0}+  \notag \\
& E^{3}\gamma _{3}\wedge \gamma _{0})-\beta \gamma (E^{2}\gamma _{2}\wedge
\gamma _{1}+E^{3}\gamma _{3}\wedge \gamma _{1}).  \label{eh}
\end{align}
(We denoted, as in Eq. (\ref{ad}), $E^{i}=F^{i0}$.) The components $(\mathbf{%
E}_{H}^{\prime })^{\mu \nu }$ ($(\mathbf{E}_{H}^{\prime })^{\nu \mu }=-(%
\mathbf{E}_{H}^{\prime })^{\mu \nu }$) can be written in a matrix form as
\begin{equation}
(\mathbf{E}_{H}^{\prime })^{\mu \nu }=\left[
\begin{array}{cccc}
0 & -E^{1} & -\gamma E^{2} & -\gamma E^{3} \\
E^{1} & 0 & \beta \gamma E^{2} & \beta \gamma E^{3} \\
\gamma E^{2} & -\beta \gamma E^{2} & 0 & 0 \\
\gamma E^{3} & -\beta \gamma E^{3} & 0 & 0
\end{array}
\right] ,  \label{mec}
\end{equation}
The same form can be easily find for $\mathbf{B}_{H}^{\prime }$ and its
components $(\mathbf{B}_{H}^{\prime })^{\mu \nu }$. (This is left for
readers.) Eq. (\ref{eh}) is the familiar form for the active LT of a
bivector, here $\mathbf{E}_{H}$, but written as a CBGQ.

(For some general bivector $N$ the components transform by the LT as the
components of a second-rank tensor
\begin{eqnarray}
N^{\prime 23} &=&N^{23},\ N^{\prime 31}=\gamma (N^{31}-\beta N^{30}),\
N^{\prime 12}=\gamma (N^{12}+\beta N^{20}),  \notag \\
N^{\prime 10} &=&N^{10},\ N^{\prime 20}=\gamma (N^{20}+\beta N^{12}),\
N^{\prime 30}=\gamma (N^{30}+\beta N^{13}).  \label{nc}
\end{eqnarray}
From (\ref{nc}) one easily find $(\mathbf{E}_{H}^{\prime })^{\mu \nu }$ (\ref
{mec}) taking into account that the components $(\mathbf{E}_{H})^{\mu \nu }$
are determined by Eq. (\ref{em}).)

It is important to note that

(i') $\mathbf{E}_{H}^{\prime }$ and $\mathbf{B}_{H}^{\prime }$, in contrast
to $\mathbf{E}_{H}$ and $\mathbf{B}_{H}$, are not parallel to $\gamma _{0}$,
i.e., both $\mathbf{E}_{H}^{\prime }\wedge \gamma _{0}\neq 0$ and $\mathbf{B}%
_{H}^{\prime }\wedge \gamma _{0}\neq 0$, and thus there are the space-space
components, $(\mathbf{E}_{H}^{\prime })^{ij}\neq 0$ and $(\mathbf{B}%
_{H}^{\prime })^{ij}\neq 0.$ Furthermore,

(ii') \emph{the components} $(\mathbf{E}_{H})^{\mu \nu }$ ($(\mathbf{B}%
_{H})^{\mu \nu }$) \emph{transform upon the active LT again to the components%
} $(\mathbf{E}_{H}^{\prime })^{\mu \nu }$ ($(\mathbf{B}_{H}^{\prime })^{\mu
\nu }$); \emph{there is no mixing of components}. \emph{Under the active LT}
$\mathbf{E}_{H}$ \emph{transforms to} $\mathbf{E}_{H}^{\prime }$ \emph{and} $%
\mathbf{B}_{H}$ \emph{to }$\mathbf{B}_{H}^{\prime }.$ Actually, as already
said, this is the way in which every bivector transforms under the active LT.

Instead of using the active LT we can deal with the passive LT. The
essential difference relative to the usual covariant picture is the presence
of the basis in a CBGQ. The existence of the basis causes that every 4D CBGQ
is invariant under the passive LT; the components transform by the LT and
the basis by the inverse LT leaving the whole 4D CBGQ unchanged. This means
that such CBGQ represents \emph{the same physical quantity }for relatively
moving 4D observers. For some general bivector $N$ the components transform
according to (\ref{nc}), whereas the basis $\gamma _{\mu }^{\prime }\wedge
\gamma _{\nu }^{\prime }$ transform by the inverse LT giving that the whole $%
N$ is unchanged
\begin{equation}
N=(1/2)N^{\mu \nu }\gamma _{\mu }\wedge \gamma _{\nu }=(1/2)N^{^{\prime }\mu
\nu }\gamma _{\mu }^{\prime }\wedge \gamma _{\nu }^{\prime },  \label{enc}
\end{equation}
where all primed quantities are the Lorentz transforms of the unprimed ones.
It can be checked by the use of (\ref{em}) and (\ref{mec}) that (\ref{enc})
holds for $\mathbf{E}_{H}$, i.e., that
\begin{equation}
\mathbf{E}_{H}=(1/2)(\mathbf{E}_{H})^{\mu \nu }\gamma _{\mu }\wedge \gamma
_{\nu }=(1/2)(\mathbf{E}_{H}^{\prime })^{\mu \nu }\gamma _{\mu }^{\prime
}\wedge \gamma _{\nu }^{\prime },  \label{n1}
\end{equation}
and the same for $\mathbf{B}_{H}$.

In addition, let us see how one can find the expression for $E_{v}$ from (%
\ref{he}) as a CBGQ in the $S^{\prime }$ frame and in the $\left\{ \gamma
_{\mu }^{\prime }\right\} $ basis. In the $S^{\prime }$ frame the
``fiducial'' observers (that are in the $S$ frame) are moving with velocity $%
v$ whose components are $v^{\prime \mu }=(\gamma c,-\gamma \beta c,0,0)$. Of
course, for the whole CBGQ $v$ it holds that $v=v^{\prime \mu }\gamma _{\mu
}^{\prime }=v^{\mu }\gamma _{\mu }$, where the components $v^{\mu }$ from $S$
are $v^{\mu }=(c,0,0,0)$. Then $E_{v}$ becomes $E_{v}=F^{\prime 10}\gamma
_{1}^{\prime }\wedge \gamma _{0}^{\prime }+\gamma ^{2}(F^{\prime 20}+\beta
F^{\prime 21})\gamma _{2}^{\prime }\wedge \gamma _{0}^{\prime }+\gamma
^{2}(F^{\prime 30}+\beta F^{\prime 31})\gamma _{3}^{\prime }\wedge \gamma
_{0}^{\prime }$ $-\beta \gamma ^{2}(F^{\prime 20}+\beta F^{\prime 21})\gamma
_{2}^{\prime }\wedge \gamma _{1}^{\prime }-\beta \gamma ^{2}(F^{\prime
30}+\beta F^{\prime 31})\gamma _{3}^{\prime }\wedge \gamma _{1}^{\prime })$
. If the components $F^{\prime \mu \nu }$ are expressed in terms of $F^{\mu
\nu }$ from $S$ using (\ref{nc}) then the same components are obtained as in
(\ref{mec}). \bigskip \bigskip

\noindent \textbf{IV. APPARENT\ TRANSFORMATIONS\ OF\ ELECTRIC\ AND\ }

\textbf{MAGNETIC\ FIELDS AS\ BIVECTORS\bigskip }

In contrast to the LT of $\mathbf{E}_{H}$ (and $\mathbf{B}_{H}$)$,$ Eqs. (%
\ref{ch}) and (\ref{eh}), it is accepted in the usual geometric algebra
formalism that $\mathbf{E}_{H}$ (and $\mathbf{B}_{H}$) do not transform as
all other multivectors transform, but that they transform as
\begin{equation}
\mathbf{E}_{H,at}^{\prime }=(1/2)[F^{\prime }-\gamma _{0}F^{\prime }\gamma
_{0}]=(F^{\prime }\cdot \gamma _{0})\gamma _{0},  \label{boc}
\end{equation}
where $F^{\prime }=RF\widetilde{R}$. (The subscript ``at'' is for AT.) It is
seen from (\ref{boc}) that only F is transformed while $\gamma _{0}$ is not
transformed. The transformation (\ref{boc}) is nothing else than the usual
transformation of the electric field that is given in Ref. 7, \textit{%
Space-Time Algebra,} Eq. (18.22), \textit{New Foundations for Classical
Mechanics,} Ch. 9, Eqs. (3.51a,b) and Ref. 8, Sec. 7.1.2, Eq. (7.33).

When (\ref{boc}) is written with CBGQs then instead of the LT (\ref{eh}) we
find the AT

\begin{align}
\mathbf{E}_{H,at}^{\prime }& =F^{\prime i0}\gamma _{i}\wedge \gamma
_{0}=E^{1}\gamma _{1}\wedge \gamma _{0}+  \notag \\
& \gamma (E^{2}-\beta cB^{3})\gamma _{2}\wedge \gamma _{0}+\gamma
(E^{3}+\beta cB^{2})\gamma _{3}\wedge \gamma _{0},  \label{es}
\end{align}
In (\ref{es}) $E^{i}=F^{i0}$ and $B^{i}=(1/2c)\varepsilon ^{kli0}F_{kl}$, as
in (\ref{ad}). When the components $(\mathbf{E}_{H,at}^{\prime })^{\mu \nu }$
($(\mathbf{E}_{H,at}^{\prime })^{\mu \nu }=\gamma ^{\nu }\cdot (\gamma ^{\mu
}\cdot \mathbf{E}_{H,at}^{\prime })$) from (\ref{es}) are written in a
matrix form they are
\begin{equation}
(\mathbf{E}_{H,at}^{\prime })^{\mu \nu }=\left[
\begin{array}{cccc}
0 & -E_{at}^{\prime 1} & -E_{at}^{\prime 2} & -E_{at}^{\prime 3} \\
E_{at}^{\prime 1}=F^{\prime 10} & 0 & 0 & 0 \\
E_{at}^{\prime 2}=F^{\prime 20} & 0 & 0 & 0 \\
E_{at}^{\prime 3}=F^{\prime 30} & 0 & 0 & 0
\end{array}
\right] ,  \label{et}
\end{equation}
where
\begin{equation}
E_{at}^{\prime 1}=E^{1},\ E_{at}^{\prime 2}=\gamma (E^{2}-\beta cB^{3}),\
E_{at}^{\prime 3}=\gamma (E^{3}+\beta cB^{2}).  \label{is}
\end{equation}
The same matrix form can be obtained for $(\mathbf{B}_{H,at}^{\prime })^{\mu
\nu }$ with
\begin{equation}
B_{at}^{\prime 1}=B^{1},\ B_{at}^{\prime 2}=\gamma (B^{2}+\beta E^{3}/c),\
B_{at}^{\prime 3}=\gamma (B^{3}-\beta E^{2}/c).  \label{ib}
\end{equation}
Observe that the transformations (\ref{is}) and (\ref{ib}) are exactly the
familiar expressions for the usual transformations of the components of the
3D $\mathbf{E}$ and $\mathbf{B}$, Ref. 4, Eq. (11.148), which are quoted in
every textbook and paper on relativistic electrodynamics from the time of
Lorentz, Poincar\'{e} and Einstein.

We see from (\ref{boc}), (\ref{es}), (\ref{et}), (\ref{is}) and (\ref{ib})
that

(i'') $\mathbf{E}_{H,at}^{\prime }$ and $\mathbf{B}_{H,at}^{\prime }$, in
the same way as $\mathbf{E}_{H}$ and $\mathbf{B}_{H}$, are parallel to $%
\gamma _{0}$, i.e., $\mathbf{E}_{H,at}^{\prime }\wedge \gamma _{0}=\mathbf{B}%
_{H,at}^{\prime }\wedge \gamma _{0}=0$, whence it again holds that the
space-space components are zero, $(\mathbf{E}_{H,at}^{\prime })^{ij}=(%
\mathbf{B}_{H,at}^{\prime })^{ij}=0.$ Furthermore, it is seen from the
relations (\ref{es}), (\ref{is}) and (\ref{ib}) that

(ii'') in contrast to the LT of $\mathbf{E}_{H}$ \emph{and} $\mathbf{B}_{H},$
Eq. (\ref{eh}), \emph{the components} $E_{at}^{\prime i}$ \emph{of the
transformed }$\mathbf{E}_{H,at}^{\prime }$ \emph{are expressed by the
mixture of components} $E^{i}$ \emph{and} $B^{i},$ \emph{and the same holds
for} $\mathbf{B}_{H,at}^{\prime }$.

In all geometric algebra formalisms, e.g., Refs. 7, 8, the AT (\ref{es}) for
$\mathbf{E}_{H,at}^{\prime }$ (and similarly for $\mathbf{B}_{H,at}^{\prime }
$) are considered to be the LT of $\mathbf{E}_{H}$ ($\mathbf{B}_{H}$).
However, contrary to the generally accepted opinion, the transformations (%
\ref{boc}), (\ref{es}), (\ref{et}), (\ref{is}) and (\ref{ib}) are not the
LT. The LT cannot transform the matrix (\ref{em}) with $(\mathbf{E}%
_{H})^{ij}=0$ to the matrix (\ref{et}) with $(\mathbf{E}_{H,at}^{\prime
})^{ij}=0$. Furthermore Eq. (\ref{enc}) is not fulfilled,
\begin{equation}
(1/2)(\mathbf{E}_{H,at}^{\prime })^{\mu \nu }\gamma _{\mu }^{\prime }\wedge
\gamma _{\nu }^{\prime }\neq (1/2)(\mathbf{E}_{H})^{\mu \nu }\gamma _{\mu
}\wedge \gamma _{\nu },  \label{n2}
\end{equation}
which means that these two quantities are not connected by the LT, and
consequently they do not refer to the same 4D quantity for relatively moving
observers. As far as relativity is concerned these quantities are not
related to one another. The fact that they are measured by two observers ($%
\gamma _{0}$ - and $\gamma _{0}^{\prime }$ - observers) does not mean that
relativity has something to do with the problem. The reason is that
observers in the $\gamma _{0}$ - frame and in the $\gamma _{0}^{\prime }$ -
frame are not looking at the same physical quantity but at two different
quantities. \emph{Every observer makes measurement on its own quantity and
such measurements are not related by the LT. }The LT of $\mathbf{E}_{H}$ are
correctly given by Eqs. (\ref{ch}), (\ref{eh}) and (\ref{mec}). Therefore we
call the transformations (\ref{boc}) and (\ref{es}) for geometric
quantities, and (\ref{is}) and (\ref{ib}) for components, the ``apparent''
transformations, the AT. The same name is introduced by Rohrlich$^{11}$ for
the Lorentz contraction; the Lorentz contracted length and the rest length
are not connected by the LT and consequently they do not refer to the same
4D quantity.

In the usual covariant approaches$^{4}$ the components of the 3D $\mathbf{E}%
^{\prime }$ and $\mathbf{B}^{\prime }$ are identified, in the same way as in
(\ref{sko1}), with six independent components of $F^{\prime \mu \nu }$, $%
E_{i}^{\prime }=F^{\prime i0}$, $B_{i}^{\prime }=(1/2c)\varepsilon
_{ikl}F_{lk}^{\prime }$. This then leads to the AT (\ref{is}) and (\ref{ib}%
). The 3D $\mathbf{E}^{\prime }$ and $\mathbf{B}^{\prime }$ as \emph{%
geometric quantities in the 3D space,} are constructed multiplying the
components $E_{i}^{\prime }$ and $B_{i}^{\prime }$ by the unit 3D vectors $%
\mathbf{i}^{\prime }$, $\mathbf{j}^{\prime }$, $\mathbf{k}^{\prime }$. The
important objections to such usual construction of $\mathbf{E}^{\prime }$
and $\mathbf{B}^{\prime }$ are the following: First, the components $%
E_{i}^{\prime }$ and $B_{i}^{\prime }$ are determined by the AT (\ref{is})
and (\ref{ib}) and not by the LT. Second, there is no transformation which
transforms the unit 3D vectors $\mathbf{i}$, $\mathbf{j}$, $\mathbf{k}$ into
the unit 3D vectors $\mathbf{i}^{\prime }$, $\mathbf{j}^{\prime }$, $\mathbf{%
k}^{\prime }$. Hence it is not true that, e.g., the 3D vector $\mathbf{E}%
^{\prime }\mathbf{=}E_{1}^{\prime }\mathbf{i}^{\prime }+E_{2}^{\prime }%
\mathbf{j}^{\prime }+E_{3}^{\prime }\mathbf{k}^{\prime }$ is obtained by the
LT from the 3D vector $\mathbf{E=}E_{1}\mathbf{i}+E_{2}\mathbf{j}+E_{3}%
\mathbf{k}$ . Cosequently the 3D vector $\mathbf{E}^{\prime }$ and $\mathbf{E%
}$ are not the same quantity for relatively moving inertial observers, $%
\mathbf{E}^{\prime }\mathbf{\neq E}$. Thus, although it is possible to
identify the components of the 3D $\mathbf{E}$ and $\mathbf{B}$ with the
components of $F$ (according to Eq. (\ref{sko1})) in an arbitrary chosen $%
\gamma _{0}$ - frame with the $\left\{ \gamma _{\mu }\right\} $ basis such
an identification is meaningless for the Lorentz transformed $F^{\prime }$%
.\bigskip \medskip

\noindent \textbf{V. CONCLUSIONS\bigskip }

The main conclusion that can be drawn from this paper, and Refs. 1-3, is
that the usual transformations of the electric and magnetic fields are not
the LT. It is believed by the whole physics community that the LT of the
matrix of components $(\mathbf{E}_{H})^{\mu \nu }$, Eq. (\ref{em}), for
which the space-space components $(\mathbf{E}_{H})^{ij}$ are zero and $(%
\mathbf{E}_{H})^{i0}=E^{i}$, transform that matrix to the matrix $(\mathbf{E}%
_{H,at}^{\prime })^{\mu \nu }$, Eq. (\ref{et}), in which again the
space-space components $(\mathbf{E}_{H,at}^{\prime })^{ij}$ are zero and the
time-space components $(\mathbf{E}_{H,at}^{\prime })^{i0}=E_{at.}^{\prime i}$
are given by the usual transformations for the components of the 3D vector $%
\mathbf{E}$, Eq. (\ref{is}); the transformed components $E_{at.}^{\prime i}$
are expressed by the mixture of $E^{i}$ and $B^{i}$ components. (This
statement is equivalent to saying that the transformations (\ref{is}) and (%
\ref{ib}) are the LT of the components of the 3D $\mathbf{E}$ and $\mathbf{B}
$.) However, according to the correct mathematical procedure, the LT of the
matrix of components $(\mathbf{E}_{H})^{\mu \nu }$, Eq. (\ref{em}),
transform that matrix to the matrix $(\mathbf{E}_{H}^{\prime })^{\mu \nu }$,
Eq. (\ref{mec}), with $(\mathbf{E}_{H}^{\prime })^{ij}\neq 0$. As seen from (%
\ref{mec}) all transformed components $(\mathbf{E}_{H}^{\prime })^{\mu \nu }$
of the electric field are determined only by three components $E^{i}$ of the
electric field; there is no mixture with three components $B^{i}$ of the
magnetic field.

It is worth noting that the whole consideration is much clearer when using
1-vectors $E$ and $B$, as in Refs. 2, 3, for the representation of the
electric and magnetic fields. Then, e.g., $E=(1/c)F\cdot v$. In the frame of
''fiducial'' observers it becomes $E=F\cdot \gamma _{0}$, $E=E^{i}\gamma
_{i}=F^{i0}\gamma _{i}$. By the active LT the electric field $E$ transforms
again to the electric field (according to (\ref{bo})) $E^{\prime }=R(F\cdot
\gamma _{0})\widetilde{R}=F^{\prime }\cdot \gamma _{0}^{\prime }$, i.e., $%
E^{\prime }=E^{\prime \mu }\gamma _{\mu }=-\beta \gamma E^{1}\gamma
_{0}+\gamma E^{1}\gamma _{1}+E^{2}\gamma _{2}+E^{3}\gamma $, which now
contains the temporal component $E^{\prime 0}=-\beta \gamma E^{1}$. This is
the way in which a 1-vector transforms. (Generally, for components, $%
E^{\prime 0}=\gamma (E^{0}-\beta E^{1})$, $E^{\prime 1}=\gamma (E^{1}-\beta
E^{0})$, $E^{\prime 2,3}=E^{2,3}$.) For the passive LT it holds that $%
E=E^{\mu }\gamma _{\mu }=E^{\prime \mu }\gamma _{\mu }^{\prime }$; $E$ is
the same quantity for relatively moving observers. On the other hand the AT (%
\ref{is}) for components are obtained taking that $E_{at}^{\prime
}=F^{\prime }\cdot \gamma _{0}$, only $F$ is transformed but not $\gamma _{0}
$, i.e., $E_{at}^{\prime }=0\gamma _{0}+E_{at}^{\prime i}\gamma _{i}$,

$E_{at}^{\prime }=E^{1}\gamma _{1}+\gamma (E^{2}-\beta cB^{3})\gamma
_{2}+\gamma (E^{3}+\beta cB^{2})\gamma _{3}$, and obviously $E$ and $%
E_{at}^{\prime }$ are not the same quantity for relatively moving observers,
$E^{i}\gamma _{i}\neq E_{at}^{\prime i}\gamma _{i}$. All the same as for
bivectors $\mathbf{E}_{H}$ and $\mathbf{B}_{H}$ but much simpler and closer
to the usual formulation with the 3D $\mathbf{E}$ and $\mathbf{B}$. However
there is already extensive literature, e.g., Refs. 7, 8, in which the
bivectors $\mathbf{E}_{H}$ and $\mathbf{B}_{H}$ are employed. Therefore, in
this paper, the elaboration of the fundamental difference between the AT and
the LT is given using bivectors and not 1-vectors.

These results will be very surprising for all physicists since we are all,
and always, taught that the transformations (\ref{is}) and (\ref{ib}) are
the LT of the components of the 3D $\mathbf{E}$ and $\mathbf{B}$. But, the
common belief is one thing and clear mathematical facts are quite different
thing. The true agreement of these new results with electrodynamic
experiments, as shown in Refs. 2, 3 and Refs. 5, 6, substantially support
the validity of the results from Refs. 1 - 3 and Refs. 5, 6. Ultimately,
these new results say that the Lorentz invariant 4D geometric quantities are
physical ones, and not, as usually accepted, the 3D geometric quantities.
\bigskip \medskip

\noindent \textbf{REFERENCES\bigskip }

\noindent $^{1}$T. Ivezi\'{c}, ``The proof that the standard transformations
of E and B are not

the Lorentz transformations'' Found. Phys. \textbf{33}, 1339-1347 (2003)%
\textbf{.}

\noindent $^{2}$T. Ivezi\'{c}, ``The difference between the standard and the
Lorentz

transformations of the electric and magnetic fields. Application to motional

EMF,'' Found. Phys. Lett. \textbf{18,} 301-324 (2005).

\noindent $^{3}$T. Ivezi\'{c}, ``The Proof that Maxwell's equations with the
3D E and B are not

covariant upon the Lorentz Transformations but upon the standard

transformations: The new Lorentz-invariant field equations,'' Found.

Phys. \textbf{35} 1585-1615 (2005).

\noindent $^{4}$J.D. Jackson, \textit{Classical Electrodynamics} (Wiley, New
York, 1977) 2nd ed.

\noindent $^{5}$T. Ivezi\'{c}, ``Axiomatic geometric formulation of
electromagnetism with only one

axiom: the field equation for the bivector field $F$ with an explanation of
the

Trouton-Noble experiment,'' Found. Phys. Lett. \textbf{18}, 401-429 (2005).

\noindent $^{6}$T. Ivezi\'{c}, ``Trouton-Noble paradox revisited,''
physics/0606176.

\noindent $^{7}$D. Hestenes, ``Spacetime physics with geometric algebra,''
\textit{Am. J Phys.} \textbf{71},

691-714 (2003); \textit{Space-Time Algebra }(Gordon \& Breach, New York,
1966);

\textit{New Foundations for Classical Mechanics }(Kluwer, Dordrecht, 1999)

2nd. ed..

\noindent $^{8}$C. Doran, and A. Lasenby, \textit{Geometric algebra for
physicists }(Cambridge

University, Cambridge, 2003).

\noindent $^{9}$A. Einstein, ``On the electrodynamics of moving bodies''
Ann. Physik. \textbf{17}, 891

(1905), tr. by W. Perrett and G.B. Jeffery, in \textit{The Principle of
Relativity}

(Dover, New York, 1952).

\noindent $^{10}$D. Hestenes and G. Sobczyk, \textit{Clifford Algebra to
Geometric Calculus}

(Reidel, Dordrecht, 1984).

\noindent $^{11}$F. Rohrlich, ``True and apparent transformations, classical
electrons, and

relativistic thermodynamics'' Nuovo Cimento\textit{\ }\textbf{B} \textbf{45}%
, 76-83 (1966).

\end{document}